\documentclass[pamm,a4paper,fleqn]{w-art}
\usepackage{times,cite,w-thm}
\usepackage[T1]{fontenc}
\usepackage[utf8]{inputenc}
\usepackage{amsfonts}
\usepackage{mathabx}

\usepackage{amsfonts}
\usepackage{amsmath}
\usepackage{amssymb}
\usepackage{graphicx}
\usepackage{color}
\usepackage{amsmath}
\usepackage{physics}
\usepackage{float}
\usepackage{mathtools}
\usepackage{algorithm}
\usepackage{algpseudocode}
\usepackage{tikz}

\usepackage[colorlinks=true, allcolors=blue]{hyperref}
\usepackage{mwe}
\usepackage{epstopdf}
\usepackage[doipre={DOI:~}]{uri}

\newcommand \be {\begin{eqnarray}}
\newcommand \ee {\end{eqnarray}}
\newcommand \ben {\begin{eqnarray}}
\newcommand \een {\end{eqnarray}}

\newcommand{\kv}{\mathbf{k}}

\newcommand \rmd {\rm d}

\begin{document}

\TitleLanguage[EN]
\title[APFC model for the HCP lattice]{Amplitude phase-field crystal model for the hexagonal close-packed lattice}

\author{\firstname{Marcello} \lastname{De Donno}\inst{1,}%
\footnote{Corresponding author: \ElectronicMail{marcello.de_donno@tu-dresden.de}} }

\author{\firstname{Marco} \lastname{Salvalaglio}\inst{1,2}%
    }
    
\address[\inst{1}]{\CountryCode[DE]Institute of Scientific Computing, TU Dresden, 01062 Dresden, Germany}
\address[\inst{2}]{\CountryCode[DE]Dresden Center for Computational Materials Science, TU Dresden, 01062 Dresden, Germany}

\begin{abstract}
The phase field crystal model allows the study of materials on atomic length and diffusive time scales. It accounts for elastic and plastic deformation in crystal lattices, including several processes such as growth, dislocation dynamics, and microstructure evolution. 
The amplitude expansion of the phase field crystal model (APFC) describes the atomic density by a small set of Fourier modes with slowly-varying amplitudes characterizing lattice deformations. This approach allows for tackling large, three-dimensional systems. However, it has been used mostly for modeling basic lattice symmetries. 
In this work, we present a coarse-grained description of the hexagonal closed-packed (HCP) lattice that supports lattice deformation and defects. It builds on recent developments of the APFC model and introduces specific modeling aspects for this crystal structure. After illustrating the general modeling framework, we show that the proposed approach allows for simulating relatively large three-dimensional HCP systems hosting complex defect networks.
\end{abstract}
\maketitle

\section{Introduction}
The hexagonal closed-packed lattice structure is found in a variety of industrial materials in different fields, including aerospace \cite{Zhao2022}, automotive \cite{Kulekci2008}, and medical \cite{brunette2001}. 
It also plays a critical role in the development of new materials, such as high-entropy alloys, in which multiple elements are combined in high concentrations, resulting in unique properties exceeding those of conventional alloys \cite{Ye2016,Zhao2016,Gao2016}. In most of these fields, mesoscale models enabling bridging-scale investigations are crucial for understanding structure-property relations and designing novel materials. 

The phase-field crystal model (PFC) \cite{Elder2002,Elder2004,Emmerich2012} was developed to study properties and phenomena of crystalline systems resolving atomic structures while inspecting relatively large (diffusive) timescales, thus filtering out fast atomic vibration. It was first proposed phenomenologically, building on the simplest energy functional minimized by periodic field, namely the energy functional (or Lyapuynov functional) underlying the Swift-Hohenberg equation, while considering  conservative dissipative dynamics of the order parameter\cite{Elder2002}. However, it can be derived as an approximation of the classical density functional theory of freezing \cite{RY79,elder2007,vanTeeffelen2009}.
The PFC model is based on a smooth order parameter $\psi$ representing a microscopic density field.
It has been largely used to describe mesoscale phenomena in crystals, including elastic and plastic deformation, dislocation dynamics, and growth \cite{Elder2004,Emmerich2012,Berry2014,Backofen14,GRANASY2019}. Even though allowing for investigation at relatively large timescales, a spatial resolution resolving density peaks at microscopic lengthscales is needed for simulations, with a typical requirement of $8^d$ spatial grid points per atomic lattice site, with $d$ the dimension \cite{ElderPRE2010}. This prevents the application of the model to the investigation of large, three-dimensional systems.

The amplitude phase-field crystal model (APFC) \cite{Goldenfeld2005,Athreya2006,SalvalaglioMSMSE2022}, a coarse-grained formulation of PFC, is a good candidate for mesoscale modeling of crystals, overcoming some limitations of the PFC model. Indeed, it tackles long (diffusive) timescales at relatively large length scales while still retaining detailed descriptions of interfaces, deformations, and defects (see, e.g., Refs.~\cite{SalvalaglioPRL2021,Jreidini2021}).
The main idea behind the APFC model is to describe the density field through the amplitude of its principal Fourier modes. Such amplitudes may encode information about both phases and lattice deformation. Real amplitudes describe relaxed bulk regions, with variations at interfaces akin to order parameters entering classical phase-field models. Complex amplitudes can account for distortions and defects of a reference configuration through their phase. Since the amplitudes do not vary significantly away from defects, efficient numerical approaches exploiting inhomogeneous real-space discretization were developed, allowing for the description of large systems \cite{AthreyaPRE2007,Bercic2018,Praetorius2019,SalvalaglioNPJ2019}.

Here, we show how the APFC model can describe HCP lattices and provide numerical simulations of HPC crystalline systems hosting three-dimensional dislocation networks. To this goal, we consider extensions of the APFC model we recently introduced in \cite{DeDonno2023}, which successfully described crystals with complex symmetry and of technological interest (e.g., the diamond lattice). The framework proposed therein allows for modeling lattices with a basis. Moreover, an extension to an arbitrary number of amplitudes corresponding to Fourier modes with different wavevector length, not described by the classical APFC model, has been proposed, enabling the modeling of complex lattice structures. In Sec.~\ref{s:model}, we briefly recall the basics of PFC and APFC modeling as well as the further extensions we consider. We then introduce the modeling of the HCP lattice. In Sec.~\ref{s:results}, we apply the proposed model. We illustrate large-scale simulations of complex defect networks in the HCP lattice, forming at the interface between a reference crystalline matrix and inclusions rotated about different directions. We finally draw our conclusions in Sec.~\ref{s:conclusions}.

\section{Model}\label{s:model}
\subsection{APFC model}

The phase-field crystal model describes crystalline systems over diffusive timescales \cite{Elder2002,Elder2004,Emmerich2012} through a smooth atomic density field $\psi(\mathbf{r})$ and its evolution.
It is based on the Swift--Hohenberg energy functional
\begin{equation}\label{eq:F_PFC}
    F_\psi=\int_{\Omega} \left[A\psi\mathcal{L}\psi + B\psi^2 + C\psi^3 + D\psi^4 \right] \rmd\mathbf{r},
\end{equation}
where $A$, $B$, $C$, $D$ are phenomenological parameters \cite{elder2007}, $\mathcal{L}=(q^2+\nabla^2)^2$ in its basic formulation, and $q$ is the length of the principal wave vectors for periodic minimizers of $F_\psi$. Extensions have been proposed to account for multiple competing length scales, namely through a multi-mode energy functional with $\mathcal{L}=\prod_m^M(q_m^2+\nabla^2)^2$ \cite{Mkhonta2013} with $q_m$ the $M$ different lengths in the reciprocal space (i.e., the \textit{modes}). It is worth mentioning that complex structures can also be modeled through convenient definitions of a correlation function \cite{xtal1}.

The amplitude PFC model is based on the approximation of $\psi(\mathbf{r})$ by a sum of plane waves with complex amplitudes $\eta_j$ \cite{Goldenfeld2005,Athreya2006,SalvalaglioMSMSE2022}: 
\begin{equation}
    \psi(\mathbf{r}) = \bar{\psi} + \sum_{n=1}^N \eta_n e^{{\rm i}\kv_n\cdot \mathbf{r}}+\text{c.c.},
    \label{eq:n_app}
\end{equation}
where $\bar{\psi}$ is the average density, set here to $0$ without loss of generality, 
$\kv_n$ are the reciprocal space vectors producing the desired lattice symmetry \cite{SalvalaglioMSMSE2022}, and c.c. refers to the complex conjugate. 
The advantage of APFC over PFC is that it relies on functions, the amplitudes $\eta_n$, that vary slowly in space, as opposed to the rapidly varying atomic density $\psi$. Therefore, a coarser mesh can be used, allowing for the simulation of larger systems. 
A free energy functional $F_\eta$ depending on $\eta_n$ can be obtained through a renormalization group approach \cite{Goldenfeld2005} or, equivalently, by substituting the amplitude expansion \eqref{eq:n_app} into the PFC free energy (\ref{eq:F_PFC}) and integrating over the unit cell under the assumption of constant amplitudes therein \cite{Athreya2006,SalvalaglioMSMSE2022}. 

To model the HCP crystal structure, we need to consider a multi-mode APFC model. Moreover, we use a formulation supporting the definition of Bravais lattices with basis, featuring modified amplitude functions $\widetilde{\eta}_n= \mathcal{B}_n\eta_n$, where $\mathcal{B}_n$ are complex constants defined as \cite{DeDonno2023}
\begin{equation}\label{eq:basis_coeff}
    \mathcal{B}_n = \sum_{j=1}^{J} e^{-\mathrm{i} \mathbf{k}_n \cdot \mathbf{R}_j},
\end{equation}
with $J$ the number of atoms in the unit cell, and $\mathbf{R}_j$ their positions. The energy functional in this framework can be written as \cite{DeDonno2023} 
\begin{equation}\label{eq:F_full}
    \begin{split}
        F_{\widetilde{\eta}}=\int_{\Omega} & \bigg[A \sum_{n=1}^N 
            (\widetilde{\eta}_n \mathcal{M}_{n}\widetilde{\eta}_n^*
            +\widetilde{\eta}_n^* \mathcal{M}_{n}\widetilde{\eta}_n)+ B\zeta_2+C\zeta_3+D\zeta_4 + f_\mathrm{h} \bigg] \rmd\mathbf{r},
    \end{split}
\end{equation}
where $\zeta_{2,3,4}$ are polynomials in the amplitudes whose terms depend on the lattice symmetry, defined as
\begin{equation}
\begin{split}
    \zeta_2&=\sum_{p,q=-N}^N\widetilde{\eta}_p\widetilde{\eta}_q \delta_{\mathbf{0},\kv_p+\kv_q}
    =2\sum_{n=1}^N |\widetilde{\eta}_n|^2,
    \\
    \zeta_3&=\sum_{p,q,r=-N}^N\widetilde{\eta}_p\widetilde{\eta}_q\widetilde{\eta}_r \delta_{\mathbf{0},\kv_p+\kv_q+{\kv_r}},
    \\
    \zeta_4&=\sum_{p,q,r,s=-N}^N\widetilde{\eta}_p\widetilde{\eta}_q\widetilde{\eta}_r\widetilde{\eta}_s \delta_{\mathbf{0},\kv_p+\kv_q+{\kv_r}+\kv_s},
    \label{eq:pol_resonance}
\end{split}
\end{equation}
with $\delta_{\mathbf{0},\mathbf{Q}}=1$ if $\mathbf{Q}=\mathbf{0}$ and 0 otherwise.
The (multimode) differential operator $\mathcal{M}_{n}$ reads
\begin{equation}\label{eq:MM}
        \mathcal{M}_{n}=(\mathcal{G}_n^2 +b_n) \Gamma_n,
\end{equation}
with $b_{n}$ a set of parameters tuning the relative strength of different modes (with $b_n = b_m$ if $|\mathbf{k}_n|=q_m$), $\Gamma_n$ a factor depending on the considered modes, defined as
\begin{equation}
\Gamma_n=\prod_{{q}_m\neq|\mathbf{k}_n|} [(q_m^2-|\mathbf{k}_n|^2)^2+b_m],
\end{equation}
and $\mathcal{G}_{n}$ the operator that enters the one-mode approximation: 
\begin{equation}
    \mathcal{G}_{n}=\nabla^2+2{\rm i}\kv_n\cdot\nabla.
\end{equation}
Indeed, $\mathcal{M}_{n}$ reduces to $\mathcal{G}_{n}$ for $M=1$ and $b_1=0$.
The free energy contains a penalty term $f_{\rm h}$ for differences in the amplitudes of the same family, enforcing the stability of lattice structures against stripe-like phases. It reads
\begin{equation}\label{eq:stabterm}
    f_{\rm h}=\sum_{m=1}^M  h_m \sum_{j,i>j}^{N} \left(|\mathcal{B}_i|^2|\eta_i|^2-|\mathcal{B}_j|^2|\eta_j|^2\right)^2,
\end{equation}
where $M$ is the number of modes, and $h_m$ is a set of parameters. Specific choices of $\mathbf{k}_n$ and $\mathcal{B}_n$ chosen to model the HCP lattice are given in Sec. \ref{s:hpclattice}, while other numerical and modeling parameters are specified in 
Sec.~\ref{s:results}.

Analogously to classical formulations, which approximate the conservative evolution of $\psi$ under the assumption of constant average density \cite{SalvalaglioMSMSE2022}, the evolution for the amplitudes reads
\begin{equation}
\frac{\partial \eta_n}{\partial t}=-|\kv_n|^2\frac{\delta F_{\eta}}{\delta \eta_n^*}.
\label{eq:detadt}
\end{equation}
This equation may be supplied with a conservative dynamic for the average density for proper modeling of phase transitions (e.g., enabling solid-liquid coexistence) and further extensions to account for advanced modeling of elastic relaxation \cite{SalvalaglioMSMSE2022}. Targeting here the stability of a specific crystalline phase, and exploiting the evolution of amplitudes to allow for relaxation of an initial condition up to the formation of dislocation network, we consider the dynamics as in Eq.~\eqref{eq:detadt} only without loss of generality.

\subsection{HCP lattice}
\label{s:hpclattice}
The HCP lattice consists of an ABAB stacking of triangular lattices. It has a diatomic basis, with positions
\begin{equation}\label{eq:basis}
    R_1 = \bigg(0,0,0\bigg),\quad
    R_2 = \frac{1}{\sqrt{3}}\bigg(0,1,\sqrt{2}\bigg).
\end{equation}
The primitive vectors of the direct lattice are
\begin{equation}
    \begin{split}
        \mathbf{a}_1 = \bigg(0,0,2\sqrt{\frac{2}{3}}\bigg),\quad
        \mathbf{a}_2 = \bigg(\frac{1}{2},-\frac{\sqrt{3}}{2},0\bigg),\quad
        \mathbf{a}_3 = \bigg(\frac{1}{2},+\frac{\sqrt{3}}{2},0\bigg),
    \end{split}
\end{equation}
from which the primitive reciprocal lattice vectors $\kv_n$ can be obtained by applying the well-known relation
$\mathbf{k}_l=2\pi (\mathbf{a}_m \times \mathbf{a}_n)/(\mathbf{a}_l\cdot(\mathbf{a}_m 
\times \mathbf{a}_n))$ and cyclic permutations of ($l$,$m$,$n$);
$\mathbf{k}_2,\,\mathbf{k}_3$ are the primitive vectors of triangular lattices lying on the $k_xk_y$ plane, and $\mathbf{k}_1$ determines the spacing along $k_z$ between these lattices. To describe all the equivalent Fourier modes in the amplitude expansion, we need an additional vector, which we take to be $\mathbf{k}_4 = -(\mathbf{k}_2+\mathbf{k}_3)$. We further discard the shortest reciprocal lattice vector $\mathbf{k}_1$ because the corresponding amplitude $\eta_1$ was found to be zero upon energy minimization in all bulk systems analyzed. The spacing along $k_z$ is then included by considering the third nearest neighbors. We are then using a two-mode approximation with reciprocal lattice vectors
\begin{equation}
    \begin{gathered}
        \mathbf{k}_1 = \sqrt{\frac{3}{2}}\pi(0,0,1),\quad
        \mathbf{k}_2 = 2\pi(1,-\frac{1}{\sqrt{3}},0),\quad
        \mathbf{k}_3 = 2\pi(-1,-\frac{1}{\sqrt{3}},0),\quad
        \mathbf{k}_4 = \frac{4\pi}{\sqrt{3}}(0,1,0),\\
        \mathbf{k}_{5,6} = \mathbf{k}_2 \pm \mathbf{k}_1,\quad
        \mathbf{k}_{7,8} = \mathbf{k}_3 \pm \mathbf{k}_1,\quad
        \mathbf{k}_{9,10} = \mathbf{k}_4\pm \mathbf{k}_1,
    \end{gathered}
\end{equation}
where $\mathbf{k}_1$ is only used to construct $\mathbf{k}_{5,\dots,10}$.
Finally, we conveniently rescale all reciprocal lattice vectors by a factor ${\sqrt{3}}/{(4\pi)}$, so that $|\mathbf{k}_{2,3,4}|^2=1$. This way, the shortest mode we consider has unitary length, a common choice in the classic APFC model. It produces a rescaling factor for real-space coordinates of ${4\pi}/{\sqrt{3}}$.

By introducing the basis (\ref{eq:basis}) into the definition of the coefficients $\mathcal{B}_n$, the reconstructed density field from Eq.~\eqref{eq:n_app} with amplitudes minimizing the free energy $F_\eta$ is never found to have HCP symmetry for any of the inspected parameter ranges. In other terms, the amplitude expansion so constructed is not found to be a global minimum of the considered free energy and, thus, a stable phase. We therefore construct a different expression for the atomic density, which still has maxima in the HCP lattice sites and can be recast in the APFC formulation reported above:
\begin{equation}\label{eq:hcp_density}
    \begin{split}
        \psi(\mathbf{r}) = 2\mathrm{Re}\sum_{n=2}^4 
        (1+e^{i\mathbf{k}_n\cdot\mathbf{R_2}}) e^{i\mathbf{k}_n\cdot\mathbf{r}} + 
        (1-e^{i\mathbf{k}_n\cdot\mathbf{R_2}}) e^{i(\mathbf{k}_n+\mathbf{k}_1)\cdot\mathbf{r}} + 
        (1-e^{i\mathbf{k}_n\cdot\mathbf{R_2}}) e^{i(\mathbf{k}_n-\mathbf{k}_1)\cdot\mathbf{r}},
    \end{split}
\end{equation}
from which we can define the basis coefficients
\begin{equation}\label{eq:basis_hcp}
    \begin{split}
        \mathcal{B}_j &= 
            \begin{cases}
                   1+e^{i\mathbf{k}_j\cdot\mathbf{R_2}}      & \text{if } 2\leq j\leq 4,\\
                   1-e^{i\mathbf{k}_{j-3}\cdot\mathbf{R_2}}  & \text{if } 5\leq j \leq 7,\\
                   1-e^{i\mathbf{k}_{j-6}\cdot\mathbf{R_2}}  & \text{if } 8\leq j \leq 10,
            \end{cases}
            \qquad\to\qquad
            \mathcal{B}_{2,3,4} = \frac{1}{2} - \mathrm{i}\frac{\sqrt{3}}{2}, \quad 
            \mathcal{B}_{5...10} = \frac{3}{2} + \mathrm{i}\frac{\sqrt{3}}{2}.
    \end{split}
\end{equation}
This approximation allows us to describe an HCP lattice as the global minimum of the free energy for some parameter ranges. In the following section, we select a set of parameters when this occurs and illustrate different dislocation networks in three dimensions.

\section{Numerical Examples}\label{s:results}

\begin{figure}
    \centering
    \includegraphics[width=\linewidth]{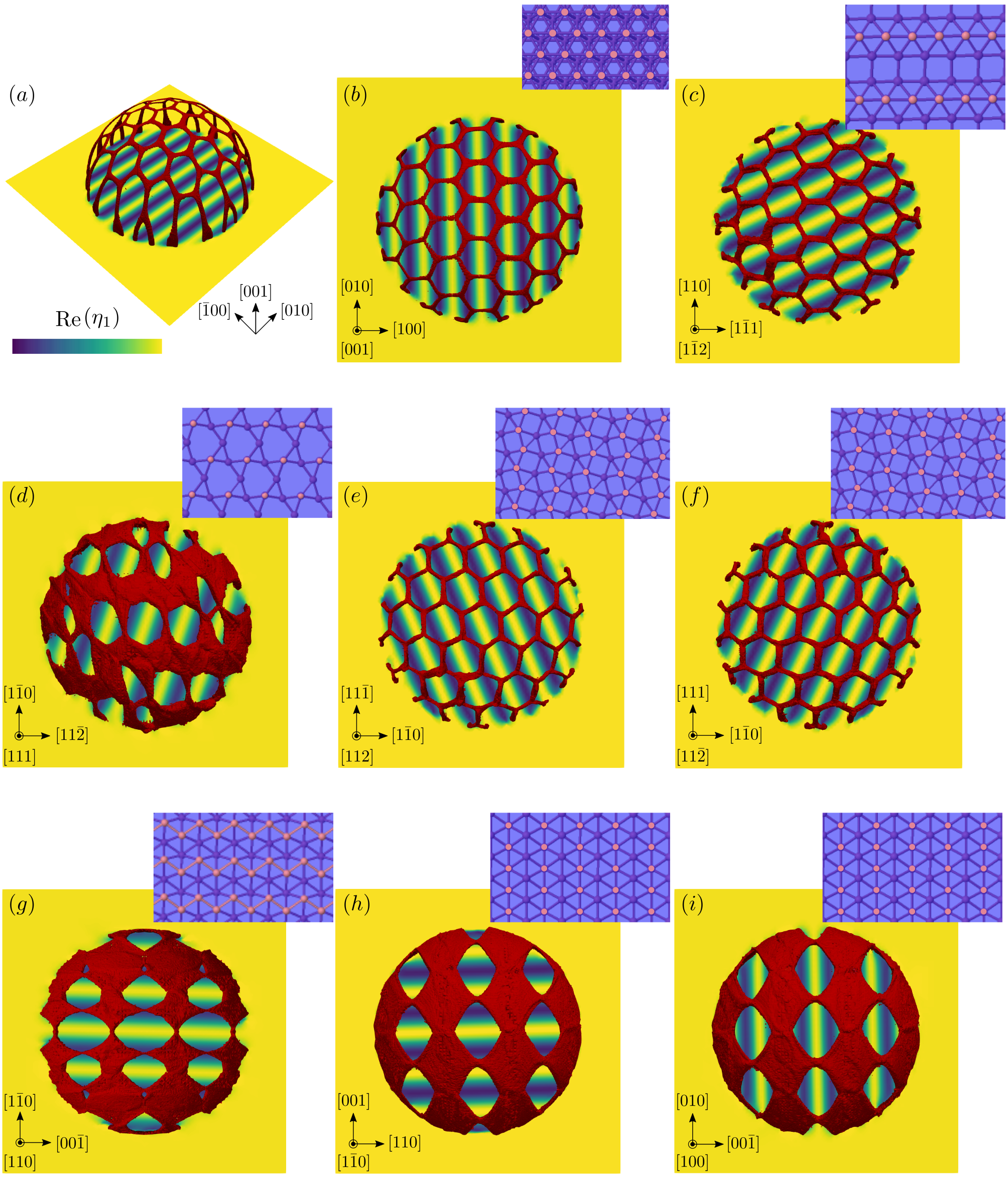}
    \caption{Simulation results for a circular inclusion rotated by a small angle $\theta=3^\circ$ about different axes. 
    The defect network is colored red, defined as the region where $\prod_n |\eta_n|^2<10^{-5}$. Re($\eta_1$) is plotted in the plane $z=0$. Insets show the atomic structure in the plane $z=0$. Atoms in the plane are pink, and atoms in successive planes are dark blue.
    Panel (a) shows the same result as (b) at a different angle to better appreciate the shape of the defect network. }
    \label{fig:inclusions}
\end{figure}

We demonstrate the capabilities of the model by showcasing relatively large three-dimensional systems. We consider a spherical inclusion enclosed in a bulk matrix and rotated by a small angle $\theta=3^\circ$ about several different crystallographic directions \cite{Salvalaglio2018}. The radius of the inclusion is set to 60 lattice spacings, and the matrix is a cube with a side of 166 lattice spacings. In terms of physical units, we may consider the equivalent size of representative HCP materials. For instance, HCP Ti has a relaxed lattice parameter $a\approx 0.295$~nm at 300~K \cite{codata2018}, meaning that for this material the side of the simulation box has a length $L\approx 49.05$~nm, and the sphere a radius $r\approx17.37$~nm. 
Simulations are performed using the adaptive Finite Element Method toolbox AMDiS \cite{Vey2007,WitkowskiACM2015}. The approach described in Refs.~\cite{Praetorius2019} allows us to solve the APFC equations efficiently with optimized mesh refinement. The parameters used are $A=0.15,\,B=-0.01,\,C=0.5,\,D=1/3,\,h_1=0,\,h_2=0.25,\,b_1=0,\,b_2=-0.1$. 

In Fig.~\ref{fig:inclusions}, we show the simulation results corresponding to the relaxation of the initial condition until the formation of the defect networks. 
We plot Re($\eta_1$) in the $z=0$ plane in each panel. This quantity is constant in the bulk matrix and oscillates slowly in the inclusion. We characterize the defect network by defining $\Lambda=\prod_n |\eta_n|^2$, which is zero at the defect since some amplitudes vanish at the dislocation cores owing to singular phases \cite{SalvalaglioMSMSE2022}. We illustrate the defects as red networks corresponding to the region where $\Lambda<10^{-5}$. For the sake of completeness and to point at the underlying crystal structure, we also show insets with the correspondingly oriented HCP atomic structure in the plane $z=0$; here, atoms in the plane are colored pink, and atoms belonging to the two adjacent planes are dark blue. The insets were realized using Jmol \cite{jmol}. 
Figure \ref{fig:inclusions}(a,b,c,e,f) show that for the orientations $z=[001],\,[1\bar{1}2],\,[112],\,[11\bar{2}]$ a rather well-defined hexagonal network forms in the region of the network extending on planes perpendicular to the rotation axis. They may be ascribed to the HCP primary slip systems\footnote{Discussions of crystallography and dislocation analysis in the HCP lattice are conventionally reported using a four-indexes corresponding to four axes: the three basal axes of the unit cell, $\mathbf{a}_1$, $\mathbf{a}_2$, and $\mathbf{a}_3$, which are separated by $120^\circ$; and the vertical axis (usually referred to as $c$ axis).} (e.g. (0001)[11$\bar{2}$0] or (10$\bar{1}$0)[11$\bar{2}$0] \cite{moffatt1967structure}). In the considered mesoscale method, we can describe the connections to such structures, peculiar of twist small-angle grain boundaries, with defects forming on other interface orientations (here connecting around the sphere chosen as the initial condition for the inclusion), c.f. Fig.~\ref{fig:inclusions}(a).

We also report different orientations where less defined defect structures form, with extended regions where the loss of coherency is observed. Similar evidence has been recently discussed for twist grain boundaries in the diamond structure \cite{DeDonno2023}. We may note that the regions where crystalline order in the crystal is retained, namely the "holes" in the defect networks, exhibit distorted hexagonal shapes. A correlation emerges between the complexity of the crystal structure on the plane perpendicular to the rotation axis and the formation of such networks. We remark, however, that knowing the geometry of the individual crystal planes perpendicular to the rotation axis is not enough to predict the shape of the defect. For example, panels $(c)$ and $(i)$ have crystal planes in which atoms are aligned but produce different defect networks. This is due to the influence of the overall crystal geometry, hinted by the adjacent crystal planes we show in dark blue in the insets. These examples are here reported as numerical experiments and serve the scope of showcasing the stability of the HCP phase in the APFC model considering systems with dislocations. Further detailed discussions are beyond the scope of the current investigation and are left to future dedicated studies.

\section{Conclusions}
\label{s:conclusions}
We devised a coarse-grained description of the hexagonal closed-packed lattice by exploiting recent extensions of the amplitude phase-field crystal model. Importantly, the description of lattices with a basis introduced in \cite{DeDonno2023}, namely the definition of $\mathcal{B}_n$, Eq.~\eqref{eq:basis_coeff}, is here applied beyond its original scope. Indeed it effectively encodes a local structure and differs from setting the relative position of atoms forming a basis per Bravais lattice site.

We demonstrated the effectiveness of our description by simulating large, non-trivial defect networks arising from material inclusions rotated by many different crystallographic axes. Our formulation is designed for real-space numerical methods, and it can reach large scales by exploiting adaptive meshes \cite{Athreya2006,SalvalaglioMSMSE2022,Praetorius2019}.
Therefore, we envisage applications in large simulations of HCP systems hosting dislocations. It is worth mentioning that the HCP description can be expanded to other lattices with a similar structure, such as hexagonal diamond or wurtzite. 

Regarding numerical efficiency, we note that the equations we used for the multimode APFC model feature the same differential order (fourth) as the classical one-mode APFC. 
Therefore, our advanced description of the HCP lattice does not introduce stiffness or complexity to the numerical problem. On the other hand, the number of equations to be solved depends on the number of reciprocal lattice vectors in the amplitude expansion, meaning that a complex lattice as HCP necessarily requires a higher computational cost than simpler lattice structures, e.g., the body-centered cubic symmetry.

Finally, we remark that the model still relies on the classic APFC equations, and is therefore compatible with other additions recently proposed within the APFC framework, such as multiphase systems, or advanced treatments of elasticity
\cite{SalvalaglioMSMSE2022}.

\section*{Acknowledgements}
We thank Lucas Benoit-{}-Maréchal for fruitful discussions on modeling HPC crystals. This work was funded by the Deutsche Forschungsgemeinschaft (DFG – German Research Foundation), Grant No.~SA4032/2-1. Computing resources were provided by the Center for Information Services and High-Performance Computing (ZIH) at TU Dresden.

\bibliographystyle{pamm.bst}
\bibliography{references.bib}

\end{document}